\documentclass[twocolumn,showpacs,preprintnumbers,amsmath,amssymb]{revtex4}

\usepackage{graphicx}
\usepackage{dcolumn}
\usepackage{bm}

\begin{document}

\title{Topological insulator ribbon: Surface states and dynamical response}

\author{Lei Hao$^{1,2}$, Peter Thalmeier$^3$, and T. K. Lee$^2$}
 \address{$^1$Department of Physics, Southeast University, Nanjing 210096, China \notag \\
 $^2$Institute of Physics, Academia Sinica, Nankang, Taipei 11529, Taiwan \notag \\
 $^3$Max Planck Institute for Chemical Physics of Solids, D-01187 Dresden, Germany}

\date{\today}

\begin{abstract}
We study theoretically the distributions of charge and spin
polarization of a topological insulator ribbon, with a
realistic rectangular cross section. Due to constriction in two
lateral directions, the surface states discretize into a series
of subbands inside of the bulk band gap. The charge and spin
distribution show interesting characters which are different
from an ideal topological surface state. The effect of merging
of four different surface states into the new one as an entity
are analyzed. Optical conductivity and dynamical spin
susceptibility of the ribbon are studied. Different from a
single ideal surface, the optical response and dynamical spin
susceptibility of a ribbon do not have a clear correspondence.
The dynamical spin susceptibility could be used to identify the
more adequate model for Bi$_2$Se$_3$.
\end{abstract}


\maketitle

\section{\label{sec:Introduction}Introduction}

The topological insulators are characterized by spin polarized
helical boundary states protected by time reversal
symmetry.\cite{kane05,bernevig06,fu07l,fu07b,roy09} In three
dimensions, the Bi$_2$Se$_3$ material has attracted great
attention because of its well defined Dirac cone like surface
state dispersion.\cite{zhang09,xia09} Though perfect
two-dimensional surface states account for the most essential
properties of topological insulator, real samples also have
lateral pairs of surfaces. For strong topological insulators,
each surface carries an odd number of gapless topological
surface modes.\cite{fu07l,fu07b} When the bulk material is
constricted in two directions, we get a ribbon (or, nano-wire)
of topological insulator.\cite{peng09} The Aharonov-Bohm effect
of the Bi$_{2}$Se$_{3}$ ribbon showed interesting
$\phi_{0}=h/e$ period oscillation and a conductance maximum at
zero flux\cite{peng09}, which are in contradiction with both
diffusive and ballistic transport and are explained
theoretically in terms of impurity scattering
effects.\cite{bardarson10,zhang10} More conventionally, the
ribbon (nano-wire) of Bi$_{2}$Te$_{3}$, which is also a
topological insulator and described by a model similar to that
of Bi$_2$Se$_{3}$ \cite{zhang09}, is considered as a promising
thermoelectrical material for applications.\cite{bejenari08}
Former theoretical studies of topological insulator ribbon
usually start from a cylindrical
sample.\cite{lee09,bardarson10,zhang10,egger10} While an
axi-symmetric configuration is more convenient for analytical
analysis, a real topological ribbon usually has a rectangular
cross-section.\cite{peng09,bejenari08} We thus study the charge
and spin distributions of the surface modes of such a realistic
topological insulator ribbon.

For a $\hat{z}\cdot(\mathbf{k}\times\boldsymbol{\sigma})$ (or
equivalently, $\mathbf{k}\cdot\boldsymbol{\sigma}$) type of
effective model for the surface states, the charge current
operator is proportional to the spin operator. The optical
conductivity for the surface states, which is defined in terms
of the current-current correlation function, is correspondingly
directly related to the dynamical spin
susceptibility.\cite{raghu10} But for a topological insulator
ribbon, in which two pairs of lateral surfaces are present, the
relationship between the optical conductivity and the dynamical
spin susceptibility would in general be more complicated.
However, the results always depend uniquely on the characters
of the model used, which also determine the properties of the
surface states. The study of these responses may thus give some
criteria to discriminate among available models for the
topological insulator Bi$_2$Se$_3$.

In a previous work, two different models used in former works
for Bi$_2$Se$_3$ were pointed out.\cite{hao11} Here, we propose
that the more adequate model for the material could be
identified by measuring the dynamical spin susceptibility of a
Bi$_2$Se$_3$ ribbon. While optical conductivities are
identical, the dynamical spin susceptibilities show clear
qualitative differences between results obtained from the two
models. The results are explained in terms of selection rule
analysis based on direct numerical calculations. For optical
conductivity, qualitatively different selection rules are in
effect between transitions along the ribbon axis
($x$ direction in this work) and transitions along
other directions.

\section{models and methods}

Since we are interested mainly in the dynamical responses
related to the topological surface states, it is reasonable to
start from the continuum model for a topological insulator
describing low energy states close to the $\Gamma$ point of the
Brillouin zone (BZ).\cite{zhang09,fu09,fu10}

Formerly, we recognized two different kinds of models that
exist in the literature for bulk Bi$_2$X$_3$ (X is Se or Te)
materials.\cite{fu09,wang10,fu10,li10nphys,zhang09,liu10b1,liu10b2,lu10,shan10,li10}
Close to the $\Gamma$ point, the two models could both be
written compactly in terms of the Dirac matrices as\cite{hao11}
\begin{equation} \label{h3d0}
H(\mathbf{k})=\epsilon_{0}(\mathbf{k})I_{4\times4}+\sum\limits_{i=0}^{3}m_{i}(\mathbf{k})\Gamma_{i}.
\end{equation}
Every unit cell contains two spin and two orbital degrees of
freedom. The model is hence written in terms of 4 by 4
matrices. The two orbitals concentrate mainly on the top and
bottom (looking along the $-z$ direction) Se layer of the
various Bi$_2$Se$_3$ quintuple units, and are labeled as 1 and
2.\cite{wang10,fu10} The basis is taken as
$\psi_{\mathbf{k}}$=[$c_{1\mathbf{k}\uparrow}$,
$c_{2\mathbf{k}\uparrow}$, $c_{1\mathbf{k}\downarrow}$,
$c_{2\mathbf{k}\downarrow}$]$^{T}$, for a certain wave vector
in the three dimensional BZ. Since the first term proportional
to the unit matrix is nonessential to topological properties of
the system, it is ignored in the following analysis. The
remaining model is particle hole symmetric. Close to the
$\Gamma$ point, the four remaining coefficient functions are
$m_{i=0,
\cdots,3}(\mathbf{k})$=$\{m+\frac{3}{2}t(k_{x}^2+k_{y}^2)+t_{z}k_{z}^{2},
3tk_{x}, 3tk_{y}, 2t_{z}k_{z}\}$, in which $t>0$, $t_{z}>0$ and
$m<0$ ($|m|$ is half of the bulk band gap).\cite{hao11} In
terms of Pauli matrices $s_{i}$ ($i$=0, $\cdots$, 3) in the
spin subspace and $\sigma_{i}$ ($i$=0, $\cdots$, 3) in the
orbital subspace, the first three Dirac matrices are defined
as\cite{zhang09,wang10} $\Gamma_{0}$=$s_{0}\otimes\sigma_{1}$,
$\Gamma_{1}$=$s_{1}\otimes\sigma_{3}$,
$\Gamma_{2}$=$s_{2}\otimes\sigma_{3}$. Two choices of the last
Dirac matrix $\Gamma_{3}$ define the two different models in
literature, which are:
(I)$s_{0}\otimes\sigma_{2}$\cite{wang10,fu10,li10nphys} and
(II)$s_{3}\otimes\sigma_{3}$\cite{zhang09,liu10b1,liu10b2,lu10,shan10,li10}.

Though a full gap exists in the bulk, gapless modes reside on
the surface. For surfaces directed along different directions,
the surface states are described by different effective models.
Take for example the surface states for a sample occupying the
lower half $z$ space, the effective model for it was solved to
be\cite{hao11}
\begin{equation}
H_{eff}(\mathbf{k})=3t(k_{x}s_{x}+k_{y}s_{y}),
\end{equation}
for model I. The charge current matrix is thus
$\mathbf{j}(\mathbf{k})=-e\mathbf{v}_{\mathbf{k}}=
-e\nabla_{\mathbf{k}}H_{eff}(\mathbf{k})=-3et \mathbf{s}$.
Where the Pauli matrices $s_{x}$ and $s_{y}$ are in terms of the two basis
$(\eta_{1})_{\beta}$=$\delta_{\beta1}$ and
$(\eta_{2})_{\beta}$=$\delta_{\beta3}$. $\delta_{\alpha\beta}$
is one for $\alpha$=$\beta$ and zero otherwise. While for model
II, the effective model for the surface states of the same
system is\cite{hao11}
\begin{equation}
H_{eff}(\mathbf{k})=3t\hat{z}\cdot(\mathbf{k}\times\mathbf{s})=3t(k_{x}s_{y}-k_{y}s_{x}).
\end{equation}
The corresponding charge current is
$\mathbf{j}(\mathbf{k})$=$-3et \mathbf{s}\times\hat{z}$. The
two bases are $\eta_{1}$=$\frac{1}{\sqrt{2}}[1, -i, 0, 0]^{T}$
and $\eta_{2}$=$\frac{1}{\sqrt{2}}[0, 0, -i, 1]^{T}$. For the
surface states introduced above residing on the $xy$ surface,
the two bases have definite spin characters for both model I
and model II.

Now consider the surface states of a sample occupying the half
space $y\le 0$. Similar to surface states on the $z=0$ surface,
the possible zero energy surface states on $y=0$ are obtained
by solving a set of four coupled second order differential
equations
\begin{equation}
H(k_{x}=0,k_{y}\rightarrow -i\partial_{y},k_{z}=0)\Psi(y)=E\Psi(y),
\end{equation}
for $E=0$ together with the open boundary conditions
$\Psi(y)|_{y=0}$=$\Psi(y)|_{y=-\infty}$=0.\cite{liu10b2,lu10}
For model I, the basis for the surface states are obtained as
$\eta_{1}$=$\frac{1}{\sqrt{2}}[1, 0, 0, 1]^{T}$ and
$\eta_{2}$=$\frac{1}{\sqrt{2}}[0, 1, -1, 0]^{T}$ by solving the
above coupled differential equations. Now, both orbitals
contribute to the surface states. The corresponding effective
model turns out to be
\begin{equation}
H_{eff}(\mathbf{k})=-3tk_{x}\sigma_{x}+2t_{z}k_{z}\sigma_{y}.
\end{equation}
While for model II, the surface states are described by
an effective model as
\begin{equation}
H_{eff}(\mathbf{k})=-3tk_{x}\sigma_x+2t_{z}k_{z}\sigma_{z},
\end{equation}
with the same basis set as for model I. The two basis states now do
not have definite spin characters. On the other hand, the Dirac
cone of the surface states becomes anisotropic, which is
a manifestation of the uniaxial anisotropy of bulk states
between the $xy$ plane and the $z$ direction.
\cite{egger10}

The above effective models for topological surface states
describe infinite surfaces. A real sample is however finite in
all three directions. For a ribbon geometry realized and
studied recently, the sample could be considered as infinite
along the direction of the ribbon axis (taken as $x$) while
finite in the other two directions (taken as $y$ and
$z$).\cite{peng09} Then, while $k_{x}$ could still be taken as
a good quantum number, real space viewpoint should be adopted
for $y$ and $z$ dependent quantities. For the sake of
simplicity, we discretize the $y$ and $z$ coordinates into
square lattices. The model is thus written as
\begin{eqnarray}
H=\sum\limits_{n_y,n_z,k_x}\psi^{\dagger}_{n_{y}n_{z}}(k_{x})
\sum\limits_{i=0}^{3}m^{'}_{i}(k_{x})\Gamma_{i}\psi_{n_{y}n_{z}}(k_{x}) \notag \\
+\sum\limits_{n_y,n_z,k_x}\{\psi^{\dagger}_{n_{y}n_{z}}(k_{x})[-\Gamma_{0}-i\Gamma_{3}]t_{z}
\psi_{n_{y},n_{z}+1}(k_{x})
+H.c.\} \notag \\
+\sum\limits_{n_y,n_z,k_x}\{\psi^{\dagger}_{n_{y}n_{z}}(k_{x})[-\Gamma_{0}-i\Gamma_{2}]\frac{3}{2}t
\psi_{n_{y}+1,n_{z}}(k_{x})
+H.c.\},
\end{eqnarray}
where $m^{'}_{i=0,
\cdots,3}(\mathbf{k})$=$\{m+3t+2t_{z}+\frac{3}{2}tk^{2}_{x},
3tk_{x}, 0, 0\}$. $n_{y}$ and $n_{z}$ label the unit cells
along the $y$ and $z$ directions. The lattice constants along
the three directions are taken as length units.

The paramagnetic current operators along three directions could
be obtained by the continuity equations.\cite{hao09} They are
written as
\begin{subequations}
\begin{equation}
j^{P}_{x}=-e\sum\limits_{n_{y},n_{z},k_{x}}\psi^{\dagger}_{n_{y}n_{z}}(k_{x})(3tk_{x}\Gamma_0+3t\Gamma_1)\psi_{n_{y}n_{z}}(k_{x}),
\end{equation}
\begin{equation}
j^{P}_{y}=e\sum\limits_{n_{y},n_{z},k_{x}}[\psi^{\dagger}_{n_{y}n_{z}}(k_{x})iD_{y}\psi_{n_{y}-1,n_{z}}(k_{x})+H.c.],
\end{equation}
\begin{equation}
j^{P}_{z}=e\sum\limits_{n_{y},n_{z},k_{x}}[\psi^{\dagger}_{n_{y}n_{z}}(k_{x})iD_{z}\psi_{n_{y},n_{z}-1}(k_{x})+H.c.],
\end{equation}
\end{subequations}
where $D_{y}$=$-\frac{3}{2}t(\Gamma_{0}-i\Gamma_{2})$ and
$D_{z}$=$-t_{z}(\Gamma_{0}-i\Gamma_{3})$. In this work, we
would focus on the zero temperature response behaviors of the
system. The dynamical conductivity is obtained from the
retarded current-current correlation functions by the Kubo's
formula.\cite{hao09,mahanbook} Ignoring contribution from the
diamagnetic current (which contributes to the zero frequency
Drude weight), the expression is
\begin{equation}
\sigma_{\alpha\beta}(\omega)=\frac{i}{\omega}\Pi_{\alpha\beta}(\omega),
\end{equation}
where the retarded current-current correlation function is
defined as
\begin{eqnarray}
\Pi_{\alpha\beta}(\omega)=-\frac{i}{V}\int_{-\infty}^{+\infty}
dte^{i\omega t}\theta(t)\langle0|[j^{P}_{\alpha}(t),j^{P}_{\beta}(0)]|0\rangle   \notag \\
=\frac{1}{V}\sum\limits_{n}[\frac{\langle0|j^{P}_{\alpha}|n\rangle\langle n|j^{P}_{\beta}|0\rangle}
{\omega+E_{0}-E_{n}+i\eta}-\frac{\langle0|j^{P}_{\beta}|n\rangle\langle n|j^{P}_{\alpha}|0\rangle}
{\omega+E_{n}-E_{0}+i\eta}]
\end{eqnarray}
$\{|n\rangle\}$ is a complete set of basis states, with
$|0\rangle$ denoting the ground state. $E_n$ is energy of the
state $|n\rangle$. $\eta$ is the positive infinitesimal and
taken as a small positive number in realistic calculations.

The spin operator of the system is defined as $\mathbf{S}$
=$\frac{1}{2}\sum_{n_{y}n_{z}}\sum_{k_{x}}\psi^{\dagger}_{n_{y}n_{z}}(k_{x})\mathbf{s}\otimes\sigma_{0}\psi_{n_{y}n_{z}}(k_{x})$.
The zero temperature dynamical spin susceptibility (for
$\mathbf{q}=0$) is thus defined as
\begin{eqnarray}
\chi_{ij}(\omega)=-\frac{i}{V}\int_{-\infty}^{+\infty}dte^{i\omega t}\theta(t)\langle 0|[S_{i}(t),S_{j}(0)]|0\rangle   \notag \\
=\frac{1}{V}\sum\limits_{n}[\frac{\langle0|S_{i}|n\rangle\langle n|S_{j}|0\rangle}
{\omega+E_{0}-E_{n}+i\eta}-\frac{\langle0|S_{j}|n\rangle\langle n|S_{i}|0\rangle}
{\omega+E_{n}-E_{0}+i\eta}].
\end{eqnarray}

\section{results and discussions}

Without loss of generality, we consider a ribbon with
$N_{y}$=$N_{z}$=20 and for $t$=$t_{z}$=0.5, $m$=-0.7. Fig. 1(a)
is the dispersion of the ribbon. Within the bulk gap region
($|E|<-m$), sub-bands form with a finite minimal gap as a
result of discretization of the topological surface states
induced by lateral constriction. As shown in Fig. 1(a), the
sub-bands are labeled symmetrically with respect to $E=0$ as
$\{$$\pm1$, $\pm2$, $\ldots$$\}$, with positive (negative) sign
denoting positive (negative) energy sub-bands and the larger
numbering representing a larger distance to $E=0$. Every state
is twofold degenerate due to combined time reversal symmetry
and inversion symmetry of the material.

\begin{figure}
\centering
\includegraphics[width=9cm,height=8cm,angle=0]{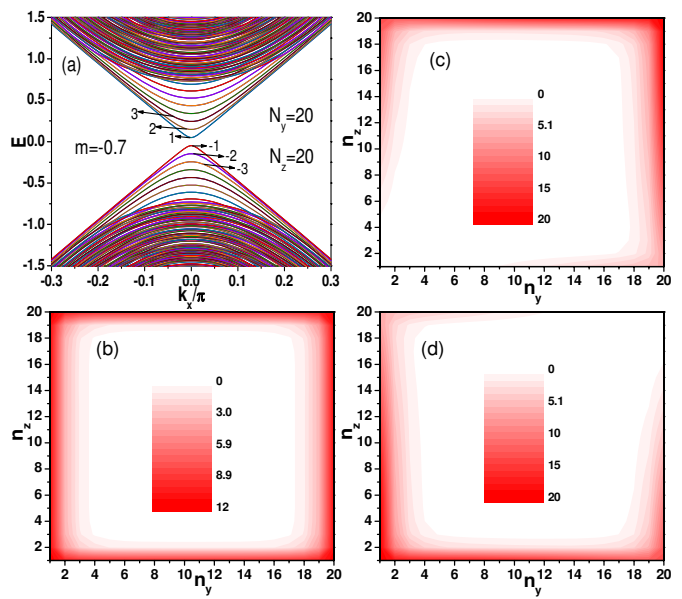}
\caption{(a) Dispersion of a topological insulator ribbon with $N_{y}$=$N_{z}$=20.
(b) Charge distribution (multiplied by $10^{3}$) of the surface modes $n$=3 for $k_{x}$=0, results for other
$n$ (for $k_{x}$=0) within the bulk gap region are similar. (c) and (d) are the charge distribution
(multiplied by $10^{3}$) of the two fold degenerate $n$=1 surface eigenmode for $k_{x}$=0.1$\pi$.}
\end{figure}

The charge and spin distributions of the topological surface
states are essential to its physical properties.  In the ribbon
geometry, an interesting feature is related to the
hybridization and redistribution of different surface states.
We give the spin and charge distributions for several typical
surface modes on the sub-bands within the bulk gap. As shown in
Fig. 1(b) for the charge distribution of a typical surface mode
for $k_{x}$=0, the charge is distributed centro-symmetrically
close to the boundary of the ribbon's cross-section. Around the
cross-section, more charge is distributed close to the four
corners which is easily understood from the smaller effective
radius there compared to the flat part of the sample. As shown
in Figs. 1(c) and 1(d), the charge distributions for the
twofold degenerate modes of $n$=1 for $k_{x}$=0.1$\pi$ are
asymmetric and centered around one corner of the ribbon,
displaying clearly the effect of hybridization of two kinds of
surface modes across the corner.

\begin{figure}
\centering
\includegraphics[width=9cm,height=10cm,angle=0]{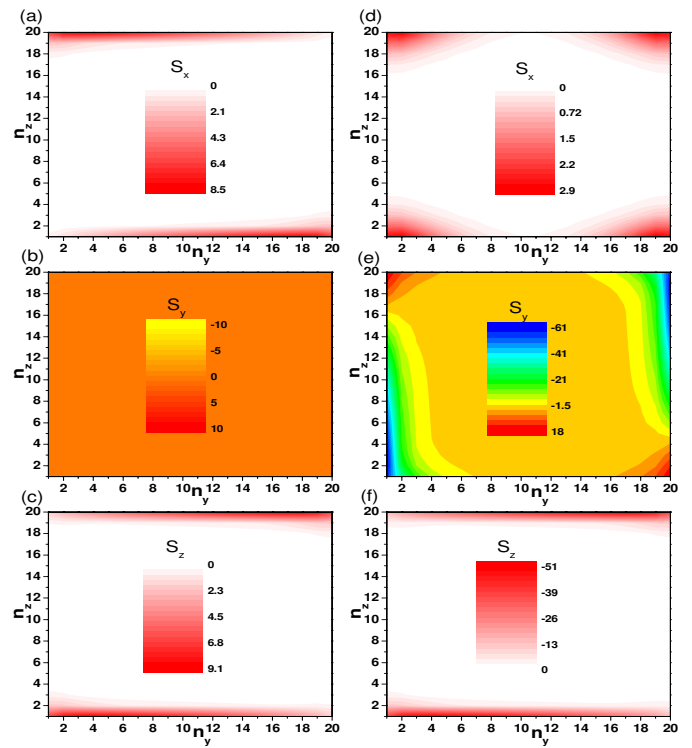}
\caption{Spin distribution for a surface mode labeled as the first sub-band at $k_{x}$=0.
The three components of the spin expectation value on every site are presented in (a), (b) and (c) (d, e and f)
for model I (model II). The results in (a), (b) and (c) (in d, e and f) are multiplied by $10^{3}$ ($10^{4}$).}
\end{figure}

While the charge distributions are qualitatively identical
along the $y$ and $z$ directions and are the same
for the two different models, the spin distribution is
more complicated. For a state residing on an infinite $xy$
surface, the spin only has in-plane component. While for states
on an infinite $xz$ surface, the basis of the surface states
indicate that the spin has both in plane and out of plane
components. In the present ribbon geometry, the surface modes
resulting from hybridization of the two types of surface modes
would be different from an ideal surface. Fig. 2 shows
distributions of the three spin components for a typical
surface mode, which is one of the two degenerate modes for
$n$=1 and $k_{x}$=0. Figs. 2(a, b, c) are for model I. A
peculiar feature is that the expectation value of the $y$ spin
component is everywhere zero. However, spin distributions of
the corresponding surface state for model II shown in Figs.
2(d, e, f) have all the three spin components nonzero along the
edge of the cross-section.

\begin{figure}
\centering
\includegraphics[width=7cm,height=12cm,angle=0]{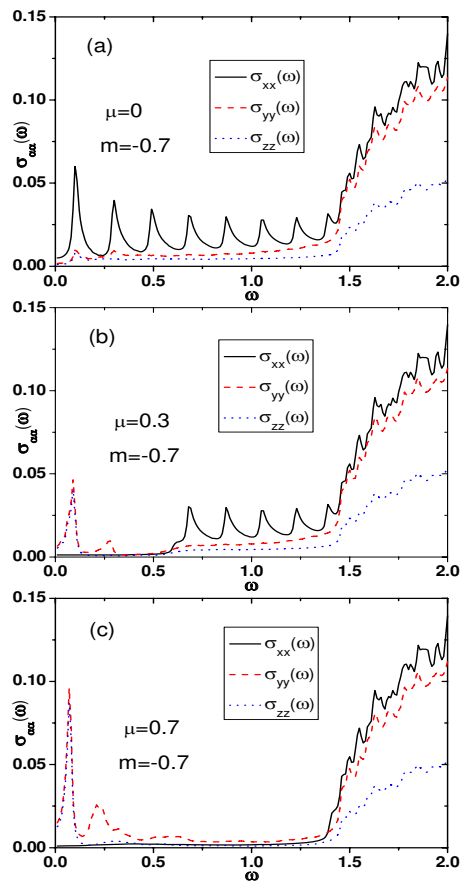}
\caption{Longitudinal dynamic conductivity (optical conductivity)
along $x$, $y$ and $z$ directions for three typical chemical potentials
(a) $\mu=0$, (b) $\mu=0.3$, and (c) $\mu=0.7$. Results for the two models
are identical. $2|m|=1.4$ is the bulk band gap. $\eta$=0.01.}
\end{figure}

Having the above charge and spin distributions of the surface
sub-bands in mind, it is natural to expect that the charge
response of the two models would be similar, but the spin
responses would possibly show very different behaviors. Fig. 3
shows the optical conductivity (in unit of $e^{2}/\hbar$) of a
ribbon with $N_{z}$=20 and $N_{y}$=20. The same results are
found for the two models. The optical conductivity is defined
as real part of the longitudinal dynamical conductivity and is
related to optical absorption. One interesting feature is the
low energy peaks of optical absorptions related to optical
transitions between the surface sub-bands inside of the bulk
band gap for $\omega<2|m|=1.4$.\cite{laforge10} When the
chemical potential lies at the charge neutral point, only the
$x$ component of the optical conductivity is significant. The
$y$ and $z$ components of the optical conductivity are smaller
and nearly featureless. When the chemical potential is shifted
away from the Dirac point by $\Delta$, the low frequency part
below 2$|\Delta|$ for $\sigma_{xx}$ is depleted while the
higher frequency part remains unchanged. In contrast, some low
frequency ($\omega$$\ll$$2|m|$) peaks emerges for $\sigma_{yy}$
and $\sigma_{zz}$. This reflects a difference in the selection
rules of optical transitions along the three directions. Since
$k_{x}$ is still a good quantum number, only vertical optical
transitions between states with the same $k_{x}$ are allowed
along all three directions, as could be seen from the current
operators defined in Eq. (8).

Differences between $\sigma_{xx}$ and optical conductivities
along the other two directions arise from selection rules with
respect to $n$ in the optical transitions. Numerically, only
transitions between pairs of sub-bands labeled by $n$ and $-n$
are allowed for optical transitions along the $x$ direction.
While for optical transitions along $y$ and $z$, the
transitions involving sub-band $n$ is largest with respect to
$n\pm1$. So, while for $\sigma_{xx}$ a series of peaks below
2$|m|$ are expected corresponding to transitions between $n$
and $-n$ sub-bands, $\sigma_{yy}$ and $\sigma_{zz}$ only have
low frequency peaks corresponding to transitions between
adjacent sub-bands. For nonzero $\Delta$, the low frequency
($\omega\le2|\Delta|$) optical transitions contributing to
$\sigma_{xx}$ are strictly forbidden since the involved
sub-band pairs are all occupied. Thus the corresponding low
frequency part of $\sigma_{xx}$ is depleted. But for
$\sigma_{yy}$ and $\sigma_{zz}$, the number of states that
could contribute to low frequency optical absorptions
continuously increases, resulting in a monotonic enhancement of
the low frequency ($\omega\le2|\Delta|$) peaks with increasing
$|\Delta|$. This nicely explains the evolution of the low
frequency optical conductivities in Fig. 3.

We now proceed to calculate the dynamical spin
susceptibilities. In accordance with the optical conductivity,
we focus on the imaginary part of the dynamical spin
susceptibilities. Only the longitudinal components would be
analyzed here. For an ideal $xy$ surface, the optical
conductivity is identical to the dynamical spin susceptibility.
In the presence of $xz$ surface, another correspondence between
optical conductivity and spin susceptibility specific to the
$xz$ surface states mixes in. In Fig. 4 we show the imaginary
parts of the longitudinal dynamical spin susceptibilities
normalized by the frequency. In agreement with expectation
based on the different spin distributions of the surface
sub-bands, the dynamical spin responses for the two different
models show qualitatively different behaviors and thus could be
used as a means to tell the right model for Bi$_{2}$Se$_{3}$.

\begin{figure}
\centering
\includegraphics[width=9cm,height=8cm,angle=0]{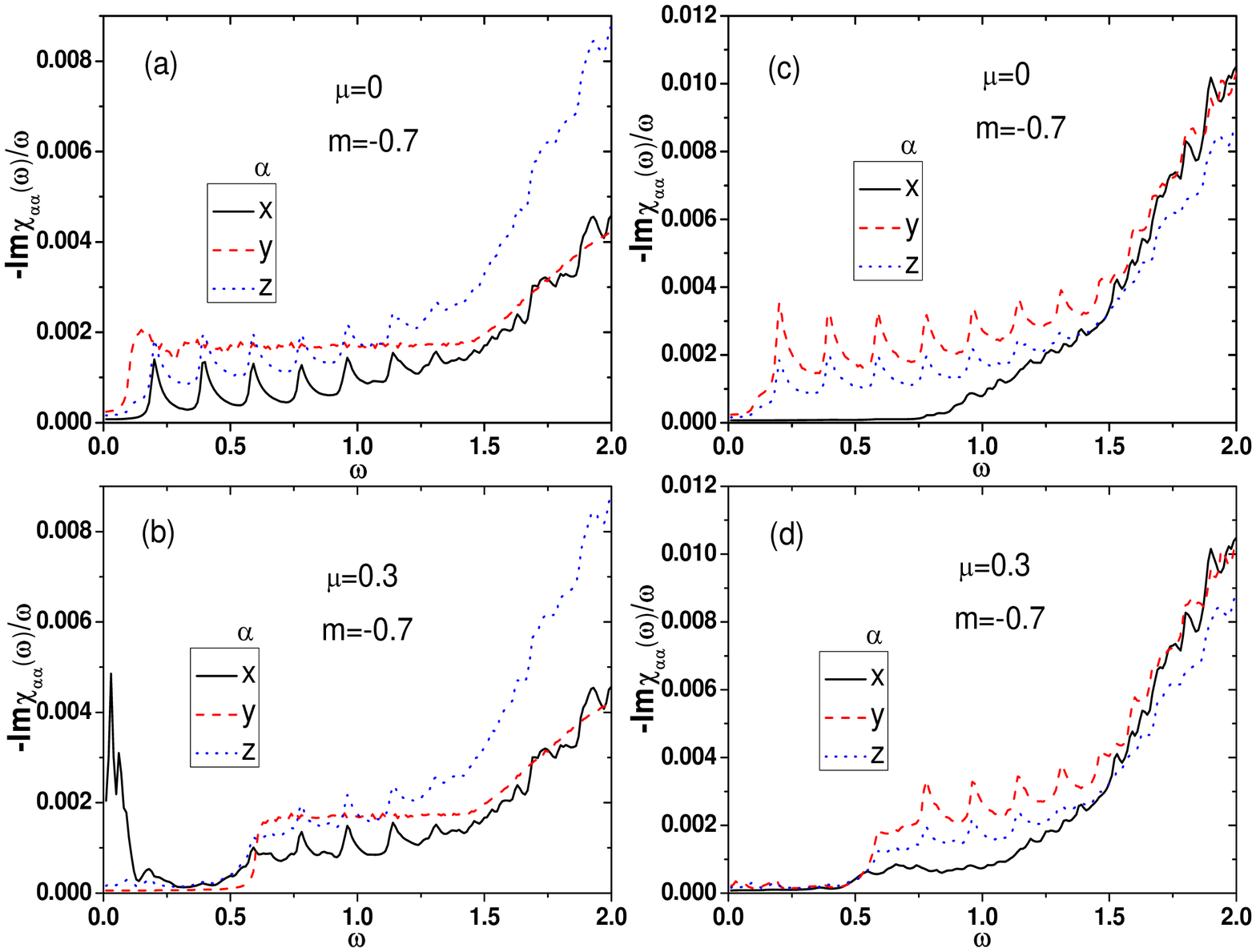}
\caption{Dynamic spin susceptibility, the imaginary part normalized by the frequency.
(a) $\mu=0$ for model I, (b) $\mu=0.3$ for model I, (c) $\mu=0$ for model II, (d) $\mu=0.3$ for model II.
$2|m|=1.4$ is the bulk band gap. $\eta$=0.01.}
\end{figure}

Though the two types of surfaces are coupled together, some
qualitative correspondences between the optical conductivity
and the dynamical spin susceptibility could still be
identified.\cite{raghu10} In particular, for an infinite $xy$
surface, the $x$ component of the optical conductivity is
proportional to the $x$ component of the dynamical spin
susceptibility for model I while it is proportional to the $y$
component of the dynamical spin susceptibility for model II.
Qualitatively similar correspondence could be identified from
comparing Fig. 4 with Fig. 3. The $y$ and $z$ components of the
optical conductivity suffer stronger influences from the
lateral constriction and the induced surface states
hybridization, the correspondences to the dynamical spin
susceptibilities are also not very straightforward. Note that,
despite the similarity with the $x$ component of the optical
conductivity, a subtle difference is that the first peak of the
dynamical spin susceptibility occurs at a frequency which is
twice that of the first peak in $\sigma_{xx}(\omega)$. This
reflects a general difference between selection rules for
electronic dipolar transitions (contributing to
$\sigma_{\alpha\alpha}(\omega)$) and magnetic dipolar
transitions (contributing to $\chi_{\alpha\alpha}(\omega)$),
which coincides only very rarely for a state such as the
surface state on the $xy$ surface. Explicitly, the matrix
elements for the spin operators are nonzero only for
transitions connecting second neighbor sub-bands, that is
between $n$ with $n\pm2$, profoundly different from the
above-mentioned selection rules for the optical conductivity.

We would like to point out an intricate point on identifying
the right microscopic model for Bi$_{2}$Se$_{3}$.
Experimentally, the spin polarization of the $xy$ surface
states was found to be perpendicular to the 2D wave
vector.\cite{hsieh09n} Though model II gives naturally the
right spin polarization, it could not directly be chosen as the
right model. Changing the spin-orbital coupling in the $xy$
plane from $\mathbf{k}\cdot\boldsymbol{\sigma}$ to
$\hat{z}\cdot(\mathbf{k}\times\boldsymbol{\sigma})$ in the bulk
model, the same spin polarization as observed in experiment is
realized also for model I.\cite{hao11,fu09,fu10} Since this
substitution amounts to a simple redefinition of the spin axes
in the $xy$ plane, $\chi_{xx}$ and $\chi_{yy}$ simply
exchange with each other and thus similar distinction between
the two models by the dynamical spin susceptibilities is still
feasible.

Finally, we mention the experimental measurements of the
dynamical spin susceptibilities. The electron magnetic
resonance (EMR) could approach $\sim$3 THz which amounts to
approximately 12 meV.\cite{hassan00} For a ribbon with much
larger cross-section than considered above but similar to
experiment,\cite{peng09} the sub-bands inside of the bulk band
gap would be much denser than for $N_{y}=N_{z}=20$. The
corresponding low frequency peaks in the dynamical spin
susceptibility would also within the reach of
EMR.\cite{hassan00} Another technique is the inelastic neutron
scattering (INS). Though it is usually hard to see surface
effects in terms of INS, when the chemical potential is tuned
close to the Dirac point inside of the bulk band gap, the
surface would possibly give the dominant signal. At last, the
spin flip surface Raman scattering could be another possible
way to measure the dynamical spin susceptibility.\cite{perez07}

\section{summary}
We have calculated the charge and spin distributions and the
corresponding dynamical spin and charge responses of a
topological insulator ribbon. Two models giving different
surface states are analyzed as a comparison. Constriction along
two lateral directions change the gapless Dirac cone like
surface states into sub-bands inside of the bulk band gap. The
corresponding charge distributions of the sub-band states
resulting from hybridization of four different surface states
are identical for the two models, giving rise to identical
optical conductivities for the two different models. The spin
distributions of the surface states are however quite different
between the two models. The dynamical spin susceptibilities are
thus also quite different and could be used to identify the
right model for Bi$_2$Se$_3$.

\begin{acknowledgments}
This work was supported by the NSC Grant No.
98-2112-M-001-017-MY3. Part of the calculations was performed
in the National Center for High-Performance Computing in
Taiwan.
\end{acknowledgments}\index{}


\end{document}